# Separating physically distinct mechanisms in complex infrared plasmonic nanostructures via machine learning enhanced electron energy loss spectroscopy

Sergei V. Kalinin[1,a], Kevin M. Roccapriore[1], Shin Hum Cho[2,3], Delia J. Milliron[2], Rama Vasudevan[1], Maxim Ziatdinov[1], and Jordan A. Hachtel[1,b]

[1] Center for Nanophase Materials Sciences, Oak Ridge National Laboratory, Oak Ridge, TN 37831

[2] McKetta Department of Chemical Engineering, The University of Texas at Austin, Austin, TX 78712

[3] Samsung Electronics, Samsung Semiconductor R&D, Hwaseong, Gyeonggi-do 18448, Republic of Korea

Low-loss electron energy loss spectroscopy (EELS) has emerged as a technique of choice for exploring the localization of plasmonic phenomena at the nanometer level, necessitating analysis of physical behaviors from 3D spectral data sets. For systems with high localization, linear unmixing methods provide an excellent basis for exploratory analysis, while in more complex systems large numbers of components are needed to accurately capture the true plasmonic response and the physical interpretability of the components becomes uncertain. Here, we explore machine learning based analysis of low-loss EELS data on heterogeneous self-assembled monolayer films of doped-semiconductor nanoparticles, which support infrared resonances. We propose a pathway for supervised analysis of EELS datasets that separate and classify regions of the films with physically distinct spectral responses. The classifications are shown to be robust, to accurately capture the common spatiospectral tropes of the complex nanostructures, and to be transferable between different datasets to allow high-throughput analysis of large areas of the sample. As such, it can be used as a basis for automated experiment workflows based on Bayesian optimization, as demonstrated on the *ex situ* data. We further demonstrate the use of non-linear autoencoders (AE) combined with clustering in the latent space of the AE yields highly reduced representations of the system response that yield insight into the relevant physics that do not depend on operator input and bias. The combination of these supervised and unsupervised tools provides complementary insight into the nanoscale plasmonic phenomena.

[a] Corresponding author, sergei2@ornl.gov
[b] Corresponding author, hachtelja@ornl.gov

The field of nanoplasmonics centers on the confinement of optical excitations to the surfaces of nanoscale structures, enabling highly controllable detection and sensing[1], enhancement of photovoltaic responses[2], novel biomedicinal applications[3], and potentially all-optical circuitry[4]. As systems become increasingly complex the need for direct nanoscale experimental investigation has made electron energy loss spectroscopy (EELS) within a scanning transmission electron microscope (STEM) a key technique due its ability to sample the true localization of the confinement at its native length scale with hyperspectral imaging[5-8]. However, the delocalized nature of plasmons results in significant overlap of distinct modes, both spatially and spectrally, often obscuring the true intensities and localizations.

As a result, there is an inherent need within plasmonic STEM-EELS analysis for exploratory data analysis methods that can both reduce the dimensionality of the EELS data sets and isolate relevant physics mechanisms that are tied to specific geometric features such as the corners of a nanoparticle, the edges of a film, etc. For core-loss EELS, the early work by Bosman[9] and others[10-13] following the visionary set of publications by Bonnett[14,15] introduced linear unmixing methods, most notably principal component analysis (PCA). Generally, linear unmixing methods represent the data as a linear combination of components (or endmembers) and corresponding loading maps (or weights). This approach reduces the dimensionality of the data set and allows visualization of regions with dissimilar behaviors. Much of the early criticisms of PCA-EELS was centered at the fact that the second and higher-order PCA components have negative values, i.e., they represented clearly non-physical behaviors since the EELS signal is formed by detector counts. The subsequent introduction of non-negative matrix factorization (NMF) and Bayesian linear unmixing[16,17] introduced the constraint of non-negativity, which is a better physical analog to EELS[16] and resulted in more physical components of the decomposition, especially for the case of plasmonics. More generally, additional physical constraints such as sum to one, spatial smoothness, and sparsity can be introduced,[18] confining the behavior of the components and loading maps to those that are more likely to be physically possible. It should be noted, however, that similar to other machine learning (ML) techniques, the components are strictly speaking defined only in a statistical sense and therefore, physical interpretation remains preponderantly qualitative unless specific physical constraints are incorporated explicitly.

Linear unmixing methods are generally characterized by two additional limitations. The first is that the spatial structure does not factor into the analysis, meaning that random transposition of the spatial pixels will not affect the component maps (unless explicit spatial regularization is used). The introduction of structured kernel Gaussian process (GP)-based analysis methods, where the interpolation specifically takes into account the spatial grid structure of the measurement[19-21] or 3D convolutional neural networks can address this issue and explore both energy and spatial domains.[22] The second limitation is that linear unmixing methods produce highly meaningful results when the physics of the system can, with a high veracity, be approximated as a sum of a number of (unknown) linear components.[23,24] This assumption holds well for the core-loss EELS signals, where the signal-generation comes from the interaction of the sub-Ångstrom electron beam with the nuclei of individual atomic species and also for simple plasmonic systems, which can be approximated straightforwardly by a small number of plasmon modes. However, for complex systems with a wide variety of modes with highly varying and often overlapping intensity, frequency, and localization, the unmixing can result in unphysical components, where weaker modes can appear only as modifications of the dominant modes making physical interpretation challenging.

Here, we propose a pathway for the supervised analysis of EELS datasets to separate and classify regions of the films with physically distinct spectral responses. The classifications are shown to be robust, to accurately capture the common spatiospectral tropes of the complex nanostructures, and to be transferable between different datasets to allow high-throughput analysis of large areas of the sample. We also demonstrate an unsupervised version of the approach using non-linear autoencoders (AE) combined with clustering in the latent space of the system AE to derive the highly reduced representations of the system response that yield insight into the relevant physics, that do not depend on operator input and bias.

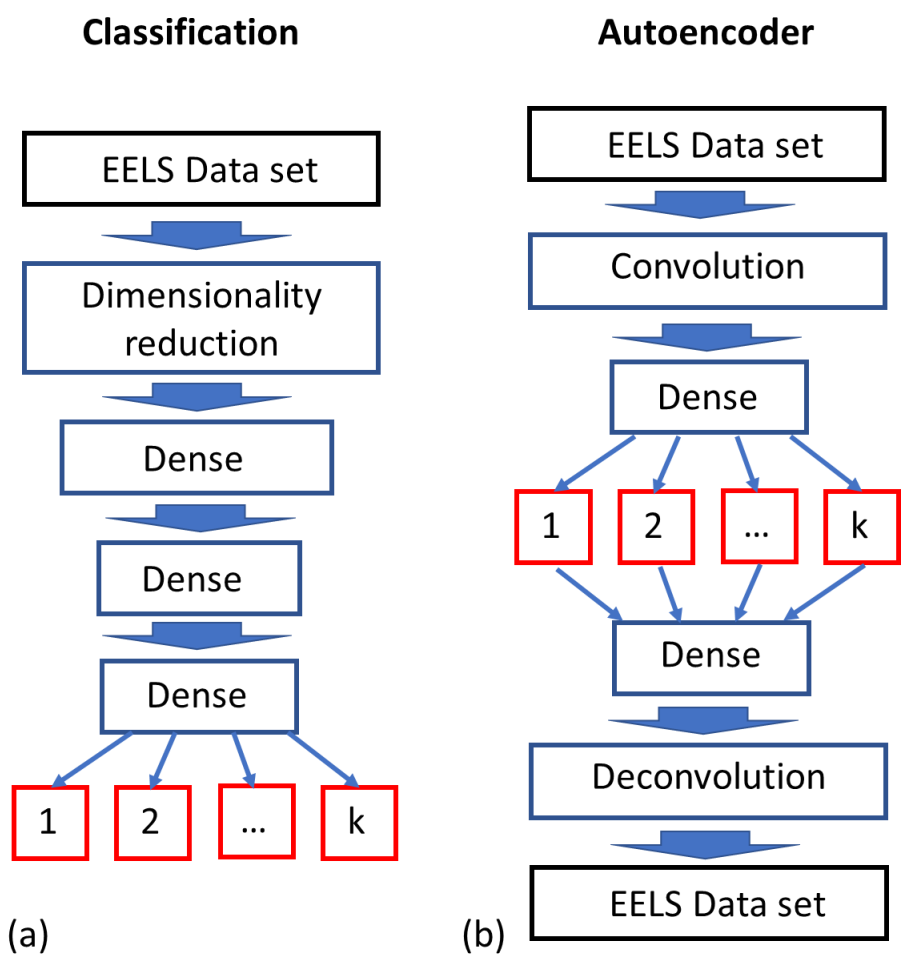

**Figure 1.** Deep neural networks-based workflows for EELS data analysis. (a) Classification of the spectral data set associates each spectrum with a specific label. Multiple variants of the network are possible, including the combination of dimensionality reduction and dense (i.e., fully connected) neural network, convolutional neural network, etc. (b) Autoencoder network compresses the data set to a small number of latent variables and subsequently expands it into original data set via a set of up-sampling and convolution or transposed convolution operations

(deconvolution). Distribution of observed behaviors in latent space provide insight into the types of materials behavior in the unsupervised manner.

The typical problem emerging in the context of the EELS data analysis is classification, i.e., interpretation of spectra at each pixel as belonging to a specific class of materials behaviors, as shown in **Figure 1** (a). Here, the network is trained using known labeled data (e.g., a subset of the image with known pixel-wise labels) and then labels can be propagated over a full data set or data obtained under similar conditions. The initial labels can be generated based on known morphological features, recognized spectral signatures, etc. Label propagation can be further applied to regions where spatial features are unavailable or where spectral images are obtained at lower spatial resolution, providing insight into the local physical mechanisms.

An alternative workflow for EELS data analysis is based on the AE approach where the pixel-wise EELS data is dimensionally reduced via a set of convolutional filters and fully connected (dense) layers into a low-dimensional latent space (bottleneck) then subsequently mapped back (deconvolve) into the original shape. The network training aims to achieve the best match between the input and output, which ideally will coincide with the bottleneck serving as the compression. The latent features can provide insight into characteristic behaviors of EELS data via spatial visualization, clustering, and density-based distance metrics, somewhat analogously to PCA and NMF analyses but occurring in the latent space of learned features. The key assumption of this analysis is that if the full dataset can be represented by a small number of latent variables then the variation of these variables across the initial dataset provides insight into the relevant physics. Note that in addition to the applications explored here, AEs are extremely powerful tools for denoising, hyperspectral image reconstruction, state compression, and other applications.[25]

For this work we examine self-assembled arrays of fluorine- and tin-doped indium oxide nanoparticles (FT:IO). These doped-semiconductor nanoparticles (NPs) are a developing class of materials that natively supports infrared plasmon resonances and are highly tunable in terms of their size, shape, and plasmonic response (e.g., localized plasmon frequency) through variations in the fluorine and/or tin doping concentration.[26,27] Moreover, while the self-assembled arrays have many interesting applications[28] they are difficult to treat theoretically due to the inherent heterogeneity of particle morphology and spacing that results from the colloidal synthesis and assembly processes.[29] Missing particles, variations in spacing, breaks in the film, and large-scale holes can all dramatically affect array properties, involving hundreds of NPs and rendering direct simulation or calculation of the plasmonic response untenable. Therefore, the combination of EELS, which can access the real plasmonic response at these nanoscale heterogeneities, and ML, which can classify and separate physical mechanisms in these complex structures, is an ideal pathway toward effective plasmonic analysis in self-assembled arrays.

The arrays of FT:IO NPs are self-assembled via the liquid-air interface method on TEM grids with ultrathin $SiN_x$ membranes (~5 nm). Another key challenge in doped semiconductor plasmonics is access to the native infrared plasmon resonances in the FT:IO particles; ergo we must use monochromation to achieve the necessary spectral resolution with EELS. The monochromated EELS measurements are performed using a Nion HERMES UltraSTEM 100 operated at an accelerating voltage of 60 kV. The monochromator on this instrument has a variable slit allowing us to reduce the absolute energy resolution for increased signal. While at maximum monochromation the instrument can achieve an energy resolution on the scale of 5-6 meV, we find plasmon linewidths in the regime of ~100 meV, so an energy resolution of ~50 meV is chosen to maximize signal without significantly broadening the observed plasmonic features. All EELS acquisitions presented here are hyperspectral datasets, or spectrum images (SI). In an SI, the beam is rastered across the region of interest and an EEL spectrum is acquired at each probe position, resulting in a 3D dataset with two spatial dimensions and one spectral dimension.

There are two critical pre-processing steps for the data: the removal of the zero-loss peak (ZLP) and removal of X-ray outliers. The ZLP is the peak in the EEL spectrum containing all elastic scattering and non-interacting (no energy loss) electrons and for thin samples it is by far the dominant feature in the EEL spectrum (typically 3-5 orders of magnitude larger than the low-loss signal); thus, small variations in the ZLP are dominant mathematically with respect to the real plasmonic peaks. This can be done in a number of ways,[30] either by simply cutting the spectral axis off at a channel above the ZLP or by performing a power law fit to the tail of the ZLP and subtracting it from each pixel in the SI. Furthermore, datasets must be cleaned to remove the small clusters of spectrometer channels with signal orders of magnitude higher than the surrounding channels that are caused by stray X-rays hitting the scintillator in the spectrometer. This is straightforwardly achieved by averaging with the surrounding unaffected pixels. Once the spectra are processed, the two spatial dimensions can be flattened into a 1D array to provide the samples for NMF[31] decomposition.

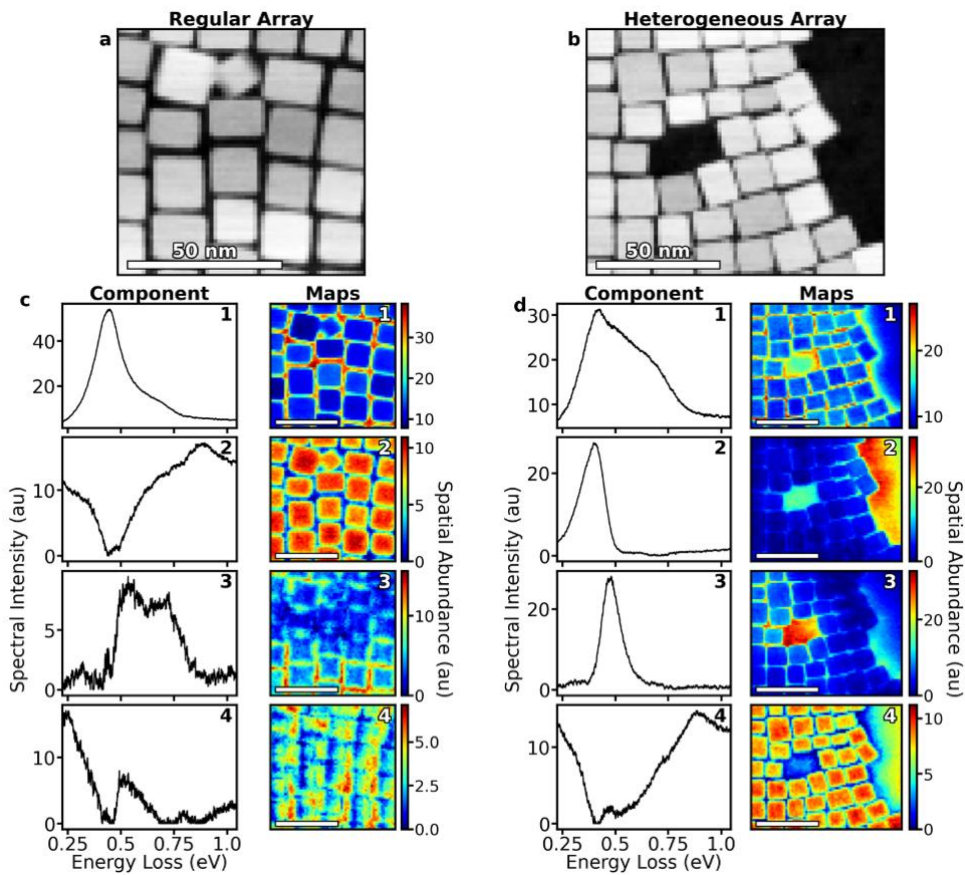

**Figure 2.** Sample NMF decompositions for FT:IO self-assembled arrays with differing degree of heterogeneity. (a,b) HAADF-STEM reference images show structure of the self-assembled film for both a regular array (RA) and a heterogenous array (HA), respectively. (c,d) 4-component NMF decompositions for the RA (c) and HA (d) datasets. The NMF decomposition can isolate heterogeneity within the samples but components are not physical and hence cannot be directly used to understand the plasmonic response in the sample. Scale bars = 50 nm.

Here, we use NMF, since the non-negative constraint offers a much better physical match to EELS compared to PCA and hence the components and abundances are more representative of the genuine features of the EELS dataset. The NMF decomposition is shown for two regions in the FT:IO films in **Figure 2**, one from a regular part of the array (referred to as Regular Array – RA) and one from a more heterogeneous part of the array (referred to as Heterogeneous Array – HA). High angle annular dark field (HAADF)-STEM reference images are shown in Fig. 2 (a,b) for the RA and HA, respectively, and show that some of the native heterogeneity in the self-assembly process can be seen in the RA image, but the HA dataset has actual missing particles in the array and is acquired at the edge of the film presenting heterogeneity with fundamentally different physical features.

We perform a relatively simplistic 4-component decomposition to demonstrate conventional NMF plasmon decomposition in hyperspectral datasets. The decomposition produces both a spectral endmember and a spatial abundance map for each component, such that the real spectrum at each pixel is approximately a linear combination of the component's spectra weighted by the component's spatial abundance at that pixel, which are shown in Fig. 2 (c,d) for the RA and HA decompositions, respectively. The NMF decomposition effectively highlights different regions of the self-assembled films and associates them with a spectrum. However, while the two datasets are acquired on the same sample, and hence one would hope for similar components, for the most part the spectral and spatial components are highly different. The only component that is similar is the one localized to the particles (component 2 of the RA and component 4 of the HA) while the others bear little similarity. This occurs because in complex systems, such as this one, multiple different phenomenon can be absorbed into single NMF components and the addition of heterogeneity in the HA dataset adds new aspects of the spectral response that are folded into the decomposition breaking the comparability of the RA and HA components. Moreover, there may be nuanced behaviors that cannot be accurately accounted for by a small number of components and are omitted entirely from the NMF decomposition. Indeed, by correlating the NMF components to the raw dataset it is observed that energy correlations can still be obtained for over 100 components (Supporting Information **Figure S1**). Thus, we can see that while NMF can produce intuitive and physical components for simple systems, in complex systems with varying degrees of heterogeneity the decomposition is unphysical and cannot be used to directly understand the true plasmonic response of the system.

However, even for complex systems, NMF is an excellent exploratory approach to guide a more rigorous analysis. For instance, we note that a large fraction of the spatial pixels of the NMF decomposition shown for the HA in Fig. 2 can be associated with specific physical features. This allows us to form an intuition for what the real physics in the system should be and we can start referring to larger scale features of the dataset with labels guided by NMF. For instance, we can see that the particles and gaps between the particles are highlighted in different NMF components and that the hole in the film and the edge of the film are highlighted in different NMF components, leading us to believe that we can understand the physics of the system by focusing on these regions individually. While the true plasmonic response of the array is more nuanced than this discrete labeling, it is of fundamental importance to classify different regions of the film according to the spectral differences (if any) observed between them. However, we need to move beyond NMF to achieve a physical description of the plasmonic response in each of these regions, which drives the desire for the ML-based approach utilized here. The goal is to classify specific pixels in the hyperspectral dataset with distinctly different plasmonic responses. Once the different regions are classified based on the ML labeling, we can average the spectra from these different regions to compare the genuine EELS data from each classification with the knowledge that they represent different physical mechanisms.

There is inherent interest in both supervised and unsupervised ML methodologies in order to make use of the hard-won expertise of researchers in the field and to side-step the biases that come with that expertise for a clear objective analysis of the data. Thus, we will perform separation of the physical mechanisms through two methods: label propagation (supervised) and latent space clustering (unsupervised).

We begin with label propagation. To achieve robust label propagation we adapt the approach for recognition imaging originally proposed for scanning probe microscopy (SPM) by Nikiforov et al.[32] a decade ago. The operator selects pixels in the hyperspectral dataset they view as representative of a specific classification; these labels are then used to train the network.

Subsequently, the trained network propagates those labels through the rest of the dataset to identify the other pixels that are most similar to the operator assigned classification. This presents an excellent use for the NMF decompositions as we can use the fact that specific physical features are highlighted with high intensity in the abundance maps to guide the creation of masks to isolate specific physical features that we have a mathematical reason to believe are fundamentally distinct. For instance, in examining the HA dataset NMF decomposition, one can simply use intensity thresholds on the NMF abundance maps to define six different labels corresponding to different physical features of the sample. It is important to note that the choice of six labels is not objective but is guided by the operator's expertise and the intuition gained from the NMF decomposition. Here, for the HA dataset we define the following six labels: particle centers, particle edges, array gaps, array holes, array edges, and the region outside the film. It is critical to avoid overlap (e.g., the same hyperspectral pixel assigned to two different labels by the operator) so the masks are chosen and refined such that each pixel has only one label.

We implemented the fully connected multilayer perceptron (MLP) with four hidden layers of 30 neurons each, *tanh* activation for internal layers and *softmax* for the output layer. The dimensionality of the input layer is determined by the number of the NMF components and was chosen to be 24. The dimensionality of the output layer is determined by the number of labels (the six labels identified in the previous paragraph). Adam optimizer for accuracy metrics and categorical cross entropy loss was used as is normal for classification problems. The network was implemented in Keras[33] with TensorFlow-2 on the back end. While the hyperparameters of the network can in principle be optimized more systematically, the accuracy was found to exceed 99.7% after the initial optimization and was assumed to be adequate for the task. We further note that a similar approach can be implemented using the convolutional networks on spectra (1D input) or local subsets of EELS data (3D convolutional inputs); however, given the volume of EELS data sets these approaches are unnecessary for the present case but can be expected to become important for larger data volumes in, for example, 4D-STEM.

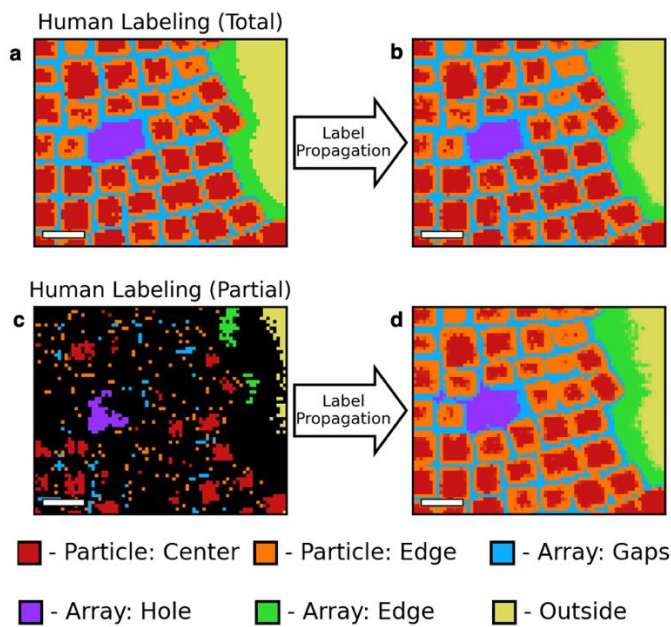

**Figure 3.** Multilayer perceptron (MLP) label propagation in complex plasmonic nanoarrays. (a) Human operator generated masks that label every pixel in an EELS hyperspectral dataset. (b) MLP propagation of labels showing little change and validating label choices as physically significant. (c,d) Human operator generated masks where majority of data is unlabeled and MLP propagation of the labels showing the same classification output as the total-human labeling dataset, demonstrating robustness of the approach. Scale bars = 20 nm.

**Figure 3** shows MLP label propagation for human-assigned classifications in the HA dataset. To start we use label propagation as a means of refining the human labeled masks. Fig. 3 (a) shows the different masks produced when the operator assigns a classification to every single pixel in the dataset based on the NMF decompositions, each color representing one operator classification. The result of the MLP reconstruction is shown in Fig. 3 (b). Here, we can see that the two are nearly identical with only a few pixels different in the refined mask. Such a result is not so surprising as the initial masks were chosen based on NMF decompositions in the first place and hence were chosen based on mathematical relevance. Nevertheless, it is encouraging to see MLP verify the human labels as viable classifications of distinct spectral behavior.

The key benefit of the MLP approach is that the total labeling shown in Fig. 3 (a,b) is not necessary for a successful label propagation process. We need only a few pixels to successfully drive the separation of the different classifications. Here, we tighten the thresholding used on the NMF abundance maps in the mask creation process to generate human labeling where only a handful of pixels is used to train the MLP and the rest are unlabeled. The partial labels are shown in Fig. 3 (c), where the black pixels represent unlabeled data in the hyperspectral dataset. We apply the same MLP label propagation process and observe that the MLP network returns an almost identical labeling of the overall dataset. There are two benefits to the match between Fig. 3 (b) and 3 (d): reduction of time spent on operator-labeling and the removal of bias from the operator labeling. In order to achieve the labeling shown in Fig. 3(a) the operator must carefully choose the thresholds to optimize the labels and in the event of overlapping labels determine which pixels go with which label, introducing a large degree of bias in the system. Since only a small number of pixels are required to achieve rigorous labeling, the operator can be less deliberate with the thresholding of the NMF components or potentially only select a few key areas manually for the initial labels (without using NMF at all). This enables more regions to be examined in the same time period, without the result being assumed in advance and without sacrificing rigor. This is critical for high-throughput analysis of specimens such as those where there may be many different regions that must be analyzed to understand the composite behavior of the mesoscale array. By further restricting the number of labeled pixels we can see that meaningful label propagation is achieved with networks trained on only 0.31% of the total dataset (Fig. S2). Only a few key representative pixels need to be chosen and the MLP label propagation can identify all other like pixels in the dataset with comparable accuracy to a human operator. This enables the ability to rigorously explore datasets where the operator's intuition only leads to being able to accurately classify a few pixels and allows the operator to experiment with different labels to try and identify the labels that most physically represent the real data.

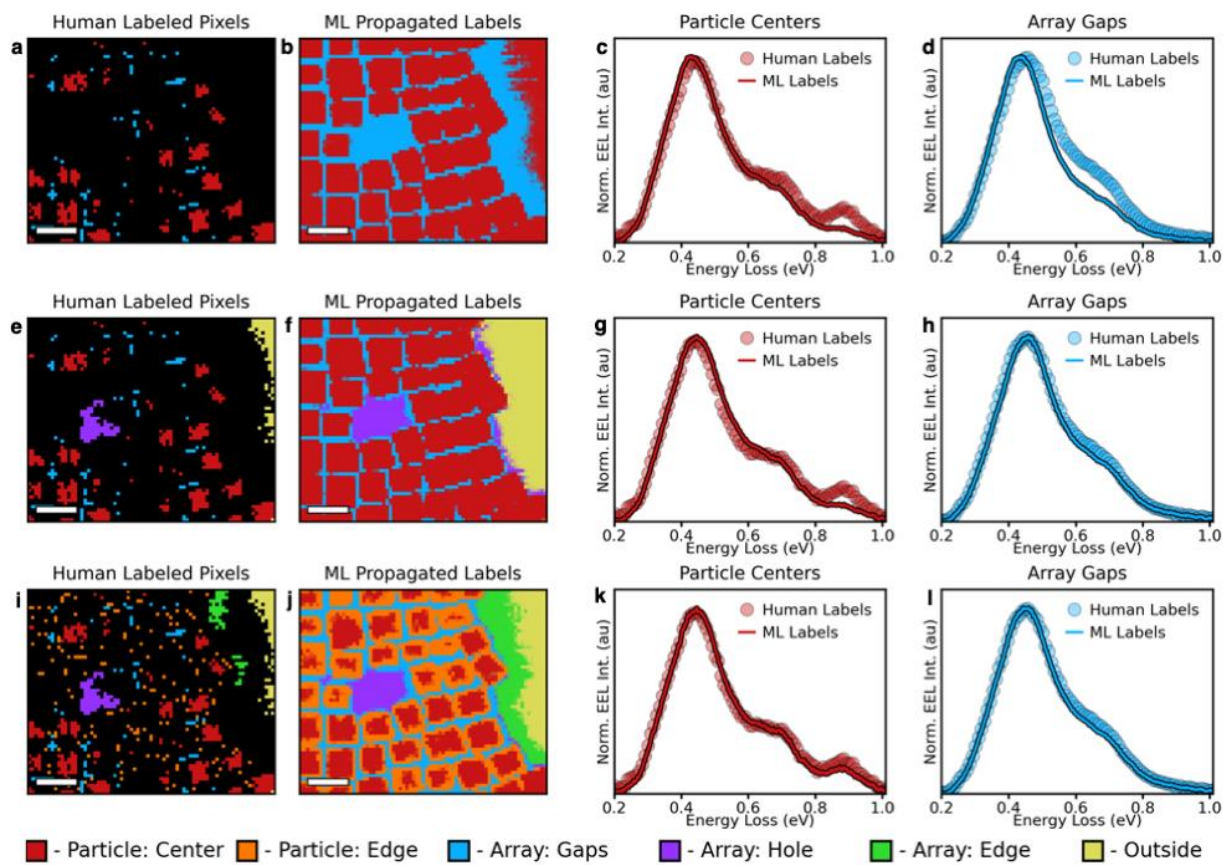

**Figure 4.** Physical accuracy of classification in bag-of-features MLP labeling. (a) The human-assigned labels for only two of the six classes identified in Fig. 3 (particle centers and array gaps) and the MLP propagated labels across the entire dataset showing how the MLP process assigns all pixels to the most valid label. (c,d) Comparison of spectra from human-labeled pixels and MLP labeled pixels for particle centers (c) and array gaps (d) showing quantitative and qualitative differences. (e-h) Same bag-of-features approach as in (a-d) except now with 'outside' and 'array hole' labels included in the MLP training showing more accurate label propagation and representative spectra. (i-l) MLP process for all six labels showing quantitatively and qualitatively accurate classification according to the spectra. Scale bars = 20 nm.

While the identification of comparable regions through label propagation is important, the critical aspect of this process is to separate different physical mechanisms based on the spectral response. Thus, the important question to ask is 'How physical are the propagated labels?' We address this question in **Figure 4** using the bag-of-features approach (where the network is only trained on a few of the operator-identified labels). By comparing the representative spectra for the propagated labels, we can see how they match the spectrum from the pixels selected by the operator. The better the match, the more accurately the MLP has identified a truly unique physical mechanism and its localization.

We first train the network on only two labels (particle centers and array gaps). The operator-labels and the MLP propagated masks are shown in Fig. 4 (a,b). Even with only two labels, MLP propagation results in a reasonably physical result; all gaps are labeled as gaps, all particles are labeled as particles. The only problem area is the outside region, which is unphysically labeled a particle. However, either label would have been a bad classification of this region, so the unphysical result is unavoidable due to insufficient labels. The importance of this unphysical labeling can be seen in the representative spectra shown in Fig. 4 (c) for the particles and 4 (d) for the gaps. Both regions show significant differences between the human-labeled spectrum and the ML-labeled spectrum. In the particle spectra, we see that the shoulder at 0.7 eV from the human-labeled dataset is suppressed in the ML dataset and the 0.9 eV peak is almost absent. In the array spectra, the 0.7 eV shoulder is also suppressed in the MLP spectrum and the dominant peak at 0.45 eV is slightly noticeably redshifted compared to the human-labeled peak. These differences indicate that each of the MLP propagated labels contain pixels exhibiting a different physical response from the pixels in the initial human label and that the propagation is unphysical.

The failure of the two-label propagation in Fig. 4 (a-d) follows necessarily from not including sufficient labels in the MLP training. We perform the same analysis including the 'outside' and 'array hole' labels in the MLP training and the results are shown in Fig. 4 (e,f). By even a cursory examination of the MLP labels we observe that a more physical result has been achieved as now the pixels outside the film are labeled as such (as opposed to particle centers) and that the hole in the center of the array is labeled as such (as opposed to an array gap). As a result we see a much better match between the representative spectra for the four-label classification in Fig. 4 (g,h). We see a reasonably strong match in the intensity of the 0.7 eV shoulder for both regions and the redshift of the ~0.45 eV peak is gone. Both improvements clearly arise from having incorporated physical mechanisms localized to the hole/edge/outside regions into labels corresponding to particle centers and array gaps. With more labels comes a better physical representation of the label in both the representative spectrum and its spatial profile in the dataset. We still see a significant difference in the particle centers at the 0.9 eV peak, so we return to the six-label classification (now including the particle edge and array edge labels used in Fig. 3) for Fig. 4 (i-l). Here, we see the reason why including separate labels for the particle center and particle edge is critical, since the 0.9 eV peak for the particle center spectrum matches quantitatively with the human-labeled data. This peak corresponds to the high energy non-radiative breathing mode that is localized to the volume of the particles. The degree of localization to the volume is significant and the mode is only excited strongly at the center of the particle and not at the particle edges and surfaces, which is consistent with depletion region effects macroscopically measured for this type of material[34]. Thus, when we exclude the edges from the particle center label, we achieve a more accurate representation of the physical mechanism and the MLP- and human-labeled data match more closely. Similarly, the array gap label is not affected by the inclusion of the array edge label because the array edge region was labeled as outside in the four-label classification; thus, the separation between the outside and the array edge does not influence the spectrum of the array gap. It is important to note that even the six-label classification is incomplete and imperfect and misses some of the complexity and nuance of the ground truth of the system, but the label-propagation approach enables robust repeatable separation of different mechanisms in a complex system without any decomposition into components, which can be unphysical or that cut-out low significance information.

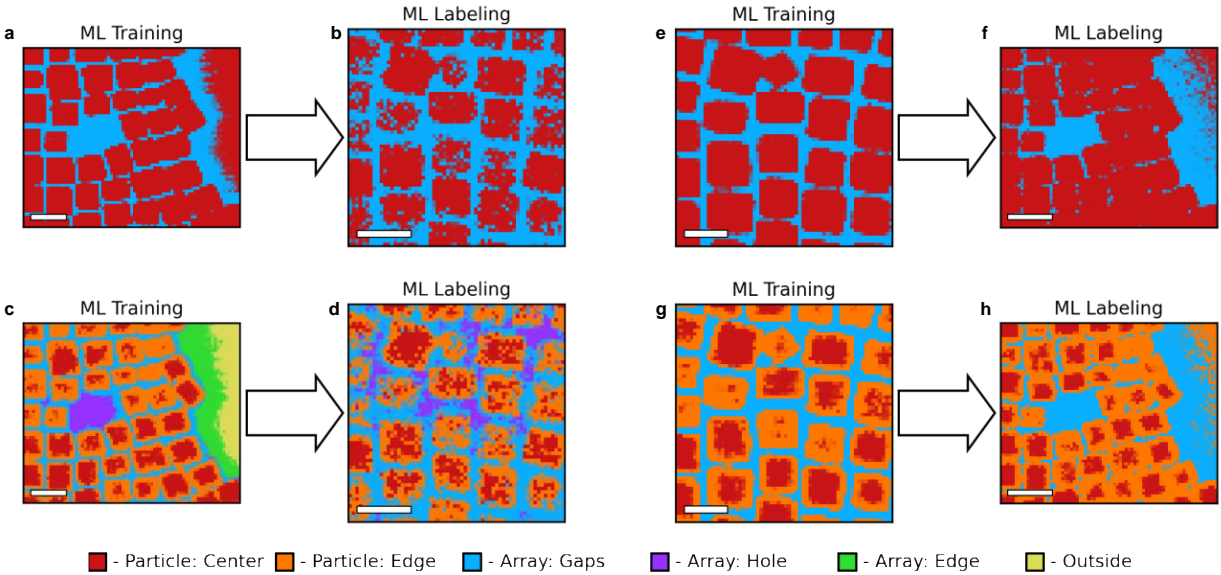

**Figure 5.** Transferability of training approach between different datasets containing comparable features. (a) MLP network is trained on the HA dataset using only two labels (particles and gaps). (b) Trained network is then applied to classify the RA dataset. (c,d) Network is trained on the full six-labels (as shown in Fig. 4) and applied to the RA dataset. (e,f) Transferability of a two-label MLP training on the RA dataset applied to the HA dataset, opposite of HA-RA transferability shown in (a,b). (g,h) Transferability of a three-label MLP training on the RA dataset applied to the HA dataset (particle centers, particle edges, array gaps). Transferability provides meaningful labels in both directions. Scale bars = 20 nm.

We can now use the trained MLP network to separate other hyperspectral datasets into the desired labels without the human operator label identification or the MLP training steps. We test the transferability of the MLP network between the HA and RA dataset and vice versa. The two datasets are from the same sample so comparable physical mechanisms should result in comparable spectral features that could be identified by the MLP network. First, we train the network only using the particle center and array gap labels on the HA dataset, the same as in **Figure 4** (a-d). We then use this network to classify the particle centers and array gaps in the RA dataset, as shown in **Figure 5** (a,b) and we observe the critical result that the MLP network can label particle centers and array gaps accurately in an entirely different dataset.

To further test the transferability, we include the other labels from the six-label classification of the HA dataset. This is a critical test, as the region has no visible array holes and no array edge or outside region, so we can test how accurately the MLP can find features even when there are labels that are not represented in the dataset. The result is shown in Fig. 5 (c,d) where it is seen that the MLP network finds the particle edges and centers effectively and divides the space between the particles between the array gap and array hole labels. While initially it can seem unphysical for the gaps in the RA dataset to be labeled as holes, it is important to note that the spacing between particles is actually greater in the areas labeled holes (at the top of the dataset) and possesses some misoriented particles while the areas labeled gaps (towards the bottom) are more regular. This indicates that the spectral response of areas with larger spacings and less regularity in the array are more physically connected to holes in the array than they are to the standard gaps of a regular array. The transfer also shows that labels that are inarguably not present in the RA (the array edge and outside labels) are not applied to any pixels in the MLP labeling. This is extremely encouraging since the difference between a hole and a gap is partly subjective (how big does a gap have to be before it becomes a hole?) but the end of the film is more tangible. The fact that no significant weight is imparted to labels that are clearly not present in the dataset validates the transferability of the MLP training between datasets.

The process is also reversible. **Figure 5** (e,f) show the two-label training from the RA dataset (particle center and array gap) applied to the HA dataset and achieves extremely similar results to the two-label classification of the HA dataset shown in Fig. 4 (a,b), where the MLP is trained directly on the HA. While it is not possible to train the MLP using all six labels from Fig. 3 and Fig. 4 on the RA dataset (which does not have the visible heterogeneities like the film edge and holes), we perform a three-label training (particle center, particle edge, and array gap) and see that the MLP-labeled pixels are the most accurate classification of the labels available and successfully distinguishes between the particle centers and particle edges based on the intensity of the high-energy breathing mode. This demonstrates the possibility of training a 'master MLP' that could be applied to any dataset with any variety of heterogeneity and have those heterogeneities accurately classified and visualized for high-throughput characterization of self-assembled films.

We note that while the MLP label propagation within a single dataset is robust, the transferability of the network to other datasets is significantly less so. An example is shown in **Figure S3**, where we perform two different types of background subtraction on the RA dataset before using the MLP network trained on the HA dataset for classification. Here, if we use the same method of background subtraction where we fit a power law to the ZLP tail and subtract it from each spectrum between the HA and the RA dataset, we return a satisfactory classification of the RA dataset, but if we use a different method (cutting the spectrum off and leaving the tail) we achieve an entirely unphysical classification. Additionally, the transferability is highly sensitive to the number of NMF components used in the initial MLP training and significant variability is observed in the transferability if only a low number are used (Fig. S4).

The alternative to a master MLP that can classify without human-operator input is a purely unsupervised approach. Here, we have adopted the convolutional AE approach, as shown in **Figure 1** (b). In the encoding stage, the signal (i.e., individual spectra) is simplified via the set of 1D convolutional filters and further translated to a fully connected latent layer. In the decoding stage, the decoder aims to reconstruct the signal from the reduced lattice representation. The encoder and decoder are trained simultaneously and jointly act to find the most efficient representation of the signal in terms of the small number of the (non-linear) latent variables. Note that a linear encoder will be similar to PCA in terms of data it provides[35].

Here, the AE architecture is implemented using five 1D convolutional layers with the *tanh* activation function and (8, 16, 32, 64, 64) convolutional filters (kernels) with a 2D latent space. The decoder was based on the set of 1D convolutional layers with *tanh* activation and upsampling with nearest-neighbor interpolation. An additional 1D convolutional layer with the ReLU activation function was added to ensure smoothness of the output and the final layer used the sigmoid function activation. The AE was implemented with the mean squared error loss and Adam optimizer with default learning rate (=0.01).

The AE training generates a manifold with an assignable number of dimensions where each coordinate on the manifold is a spectrum and the range of potential spectra is mathematically derived to represent all latent mechanisms and distributions active in the system. As a result, we can represent each pixel in our hyperspectral dataset in terms of a Cartesian coordinate on this manifold and separate physically distinct mechanisms via distance metrics in latent space without any prior knowledge. While a number of metrics and methodologies can be applied for this classification (e.g., distance-based k-means clustering or density-based DBSCAN), we utilize Gaussian mixture modeling (GMM),[36] in which a selectable number of clusters in latent space are formed based on unsupervised decomposition into Gaussian-like groups. This clustering method is naturally applicable to this case since the distributions in latent spaces are non-linear manifolds and as such, simple distance-based methods are prone to misclassification. An example of this 2D spectral manifold and how we use it to classify mechanisms in the dataset are shown in Supporting Information, Fig. S5.

We use two latent variables so the latent space can be fully visualized in the 2D plane (for higher dimensional latent spaces only marginal distributions can be visualized straightforwardly and the structure is less apparent). We also use many different numbers of GMM classifications to test the robustness of the method for an insufficient number of classifications as we did with the supervised approach shown in Fig. 4. **Figure 6** shows both the latent-space clustering and the real-space labeling from the AE with (a) two GMM classifications, (b) four GMM classifications, and (c) six GMM classifications. The results for two classifications (Fig. 6a) are extremely promising when compared to the supervised approach in Fig. 4b. We see that unsupervised labeling easily identifies the particles and the gaps, except in this example we do not achieve the unphysical result of pixels far away from the film being classified as particles (as shown in Fig. 4a). This can also be used to demonstrate the power of using advanced distance metrics in latent space, as is shown in the Supporting Information Fig. S6 where we show that k-means clustering produces the same unphysical result shown in Fig. 4b, while the GMM successfully avoids this mislabeling. The four classification labeling is also a pure improvement on the supervised approach since the AE produces much smoother labeling of the dataset than the MLP, especially in labeling he hole, which is very much off-center in Fig. 4f but is extremely symmetric with respect to the real geometry of the sample in Fig. 6b.

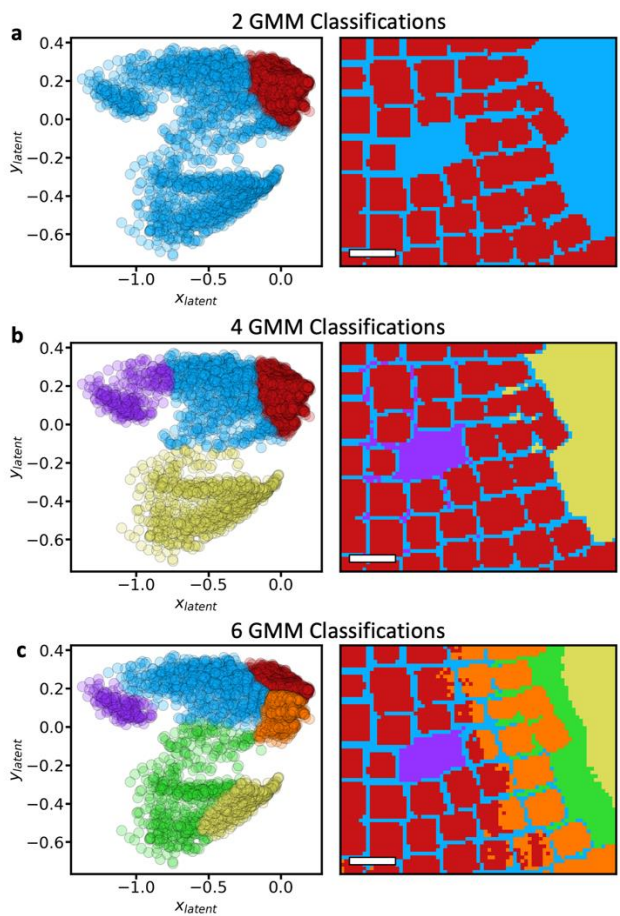

**Figure 6. Latent space classification for unsupervised labeling.** (a) GMM clustering of the 2D latent space and spatial label assignment in the HA dataset for two GMM classifications, (b) four GMM classifications, and (c) six GMM classifications showing comparable results in labeling to the operator-assisted label propagation. Scale bars = 20 nm.

Once we reach six classifications, a significant deviation between the supervised and unsupervised approach was observed. In the supervised labeling, a specific label was assigned to the particle edges to distinguish from the particle center that exhibit the breathing mode strongly, but unsupervised labeling does not return this label, instead choosing to distinguish between particles close to the edge and particles further from the edge. This difference can be understood by examining the representative spectra from the regions assigned to each label in MLP label propagation and in AE latent-space clustering.

In **Figure 7** we compare supervised MLP labeling to unsupervised AE labeling for six labels/GMM classifications. The labeling for each is shown in Fig. 7 (a,b) and show that most labels are almost identical except for the two labels associated with the particles themselves. The representative spectra from these two particle labels for MLP and AE are shown in Fig. 7 (c,d), respectively, and the other labels (array gap, array edge, array hole, and outside) are shown in Fig. 7 (e-h), respectively.

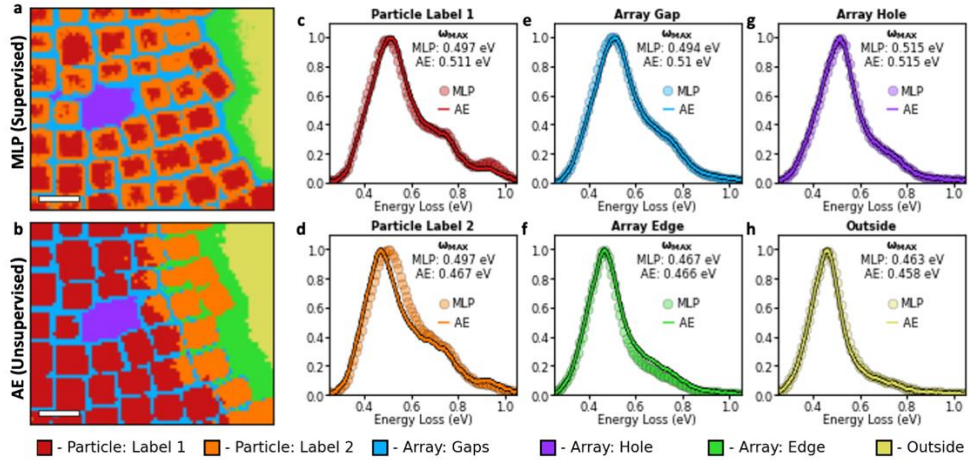

**Figure 7. Isolating different physical mechanisms with supervised and unsupervised learning.** (a) Six-label operator-assisted MLP label propagation for HA dataset and (b) six-GMM classification AE labeling for HA dataset showing similar labels for all aspects other than the two particle labels. (c-h) Comparison of representative spectra between the MLP and AE labeling: (c) Particle Label 1, (d) Particle Label 2, (e) Array Gap, (f) Array Edge, (g) Array Hole, (h) Outside regions. AE detects redshift at the edge of the array, while operator assisted MLP labels focus on localization of the breathing mode to particle centers. Scale bars = 20 nm.

In the particle labels, we note the same effect as was observed in the two- and four-label MLP classifications shown in **Figure 4**, where the 0.9 eV breathing mode peak is significantly reduced in the AE labeling compared the MLP labeling, once again due to the lack of separation between the particle edges and centers in the final labels. However, we can understand what the AE classification selected instead of the particle center/edge distinction by examining the energy of the dominant ~0.5 eV peak in each of the labels. In the operator-labeled spectra both peaks occur at the same frequency, 0.497 eV; however, in the AE the major distinction between the two labels is the frequency of this peak (0.511 eV for the first label and 0.467 eV for the second). This demonstrates that the dominant plasmon peak is significantly redshifted near the edge of the film compared with the center, which is why the AE selects the particles at the edge of the film for the second label (as opposed to the edges of the actual particles). Indeed, by looking at the frequency comparing the array gap and array edge labels we can see the same trend, even in the operator-assisted classification where the more central regions have a frequency near 0.5 eV (array gap) and the outer regions have a significantly lower frequency around 0.46 eV (array edge). This is consistent with other EELS studies on metamaterials, which found that the frequency of the dominant plasmon modes near the edge of the array were red-shifted,[37] demonstrating that this is a real physical mechanism that is selected by the AE. We also note that by using the knowledge of the redshift toward the edge of the film and choosing the initial human labeling based on proximity to the edge, the MLP label propagation returns nearly identical classifications as the AE (Fig. S7), confirming the validity of the result.

To summarize, we explored the ML workflow for identification of regions exhibiting unique physical behaviors in complex plasmonic structures. While it is established that classical linear unmixing methods are extremely useful for the initial exploratory data analysis, the high environmental dependencies of complex systems leads to large numbers of components capturing multiple degenerate aspects of the plasmonic response and become increasingly difficult to interpret. Here, we showed that recognition imaging can be used to propagate partially known labels across the data set, identifying spatial regions with a common spectral response to elucidate relevant physics across the entire sampled region. This is an excellent alternative to the unmixing approach as it involves classification of unaltered data, ensuring that no information is lost (as is the case in an NMF decomposition). This approach can be further used to establish universal networks that can extrapolate across multiple data sets and enable transition toward *ex situ* analysis of high-volume EELS datasets and automated *in situ* experimentation. We further show that such labeling can be done without the operator step by training AEs to represent an entire hyperspectral dataset in latent space such that we can effectively cluster and classify without the operator determining what they believe are the most relevant phenomena at play.

Moreover, these techniques can be universally applied for hyperspectral imaging techniques including cathodoluminescence, scanning tunneling spectroscopy, and others. We note that if the high-veracity models for the formation of the spectra are available, the networks can be trained on theoretical models and used to derive the materials parameters from experimental data.[38-40] Furthermore, conditional AEs can be used to provide unsupervised exploration of variability within known classes[41] and recent advances in the space of semi-supervised learning can be leveraged to propagate labels in diverse datasets with very sparse labels.[42-44]

This distinction observed in **Figure 7** demonstrates both the benefits and limitations of supervised and unsupervised approaches. Supervised approaches are subject to the operator's bias: if the operator believes that the particle edge/center distinction is more relevant than the array edge/center distinction, the ML-driven label propagation confirms this bias. However, while the unsupervised approach is subject to no such bias, it will necessarily favor the most mathematically relevant changes and can ignore and miss subtle yet important changes that a domain expert will not. Regardless, each methodology successfully separates, and isolates regions of the sample based on real physical mechanisms that influence the spectral response and the tandem approach allows an extremely rigorous exploration and visualization of such processes in hyperspectral data.

**ACKNOWLEDGMENTS**

This effort (electron microscopy, feature extraction) is based upon work supported by the U.S. Department of Energy (DOE), Office of Science, Basic Energy Sciences (BES), Materials Sciences and Engineering Division (K.M.R., S.V.K.) and was performed and partially supported (J.A.H., M.Z., R. K. V.) at the Oak Ridge National Laboratory's Center for Nanophase Materials Sciences (CNMS), a U.S. Department of Energy, Office of Science User Facility. S.H.C and D.J.M. acknowledge (NSF CHE-1905263, and CDCM, an NSF MRSEC DMR-1720595), the Welch Foundation (F-1848), and the Fulbright Program (IIE-15151071). Electron microscopy was performed using instrumentation within ORNL's Materials Characterization Core provided by UT-Battelle, LLC, under Contract No. DE-AC05- 00OR22725 with the DOE and sponsored

**Supporting Information**

**Data and materials availability**: The codes used in this work are available as a Jupyter notebook at https://git.io/JURFs

**Materials**: Indium (III) acetate (In(ac)$_3$, 99.99%), Tin (IV) acetate (Sn(ac)$_4$), Oleic acid (OA, 90%, technical grade), Oleyl alcohol (OlAl, 85%, technical grade) were purchased from Sigma-Aldrich. Tin (IV) fluoride (SnF$_4$, 99%) was purchased from Alfa Aesar, Hexane (99.9%), Isopropyl alcohol (99.5%, Certified ACS), were purchased from Fisher Chemical. All chemicals were used as received without any further purification.

**Fluorine Doped Indium Tin Oxide (FT:IO, F,Sn:In$_2$O$_3$) Nanoparticle Synthesis**: All synthesis procedures are undertaken by employing standard Schlenk line techniques using a modification of previously reported methods for continuous slow injection synthesis of indium oxide nanoparticles[27]. 29,46 In(ac)$_3$ 1342.97 mg (4.6 mmol), SnF$_4$ 48.68 mg (5%, 0.25 mmol), Sn(ac)$_4$ 53.23 mg (3%, 0.15 mol), and oleic acid (10 ml) are loaded in a three-neck round-bottom flask in a N$_2$-filled glove box. The precursors are stirred with a magnetic bar at 600 rpm and degassed under vacuum at 120 °C for 15 min. The injection solution is added at rate of 0.2 ml/min, into 13 ml of oleyl alcohol maintained at 290 °C vented with a 19-gauge needle under inert N$_2$ gas flow. The reaction mixture turns blue a few minutes into the injection. Subsequently, growth is terminated by removal of the heating mantle and cooled by blowing air on the three-neck flask vessel. The nanoparticles are dispersed in hexane, then isopropyl alcohol antisolvent is added and the mixture is centrifuged at 7500 rpm for 10 min. The washing procedure is repeated 3 times and the nanoparticles are redispersed in 10 ml of hexane. The resultant nanoparticle dispersion is centrifuged at 2000 rpm for 3 min to remove non-dispersible aggregates and the supernatant is collected as the nanoparticle stock sample. A concentration series (0 and 10 % Sn(ac)$_4$) of doped F,Sn:In$_2$O$_3$ nanoparticles was synthesized by controlling the Sn(ac)$_4$ to In(ac)$_3$ molar precursor ratio, while SnF$_4$ was maintained at 5% molar ratio, keeping other reaction parameters identical. All data here acquired from the 0% Sn(ac)$_4$ samples.

**STEM Characterization**: A Nion aberration-corrected ultrahigh-energy resolution monochromated EELS-STEM (HERMES$^{TM}$)[45] operated at 60 kV, equipped with a Nion Iris spectrometer. The variable monochromator slit is set to allow an energy resolution of ~30-50 meV to achieve an optimal balance of energy resolution and beam current. For these experiments, a probe current of ~20 pA with a convergence angle of 30 mrad, a collection angle of 25 mrad in the EEL spectrometer, were used.

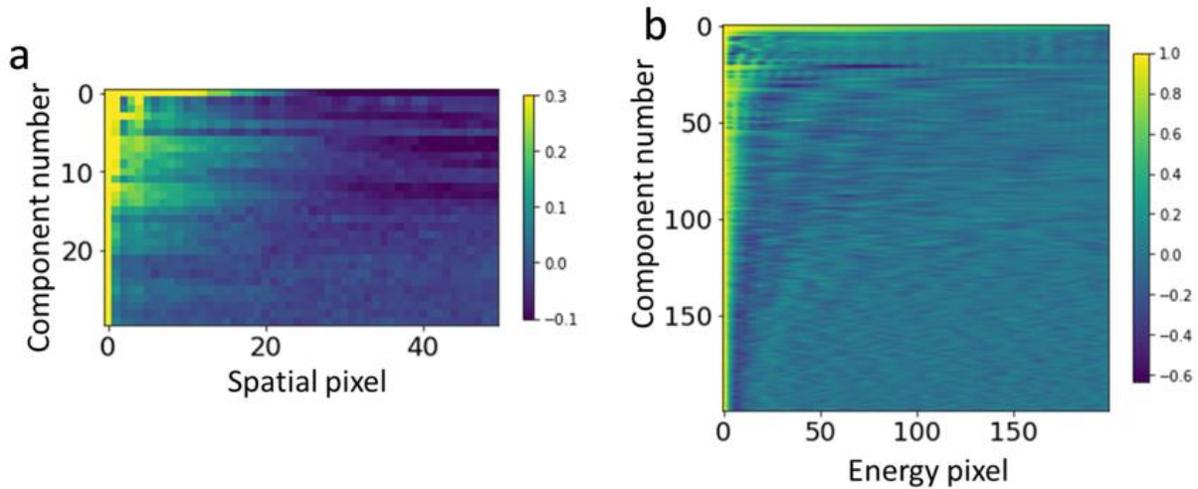

**Figure S1.** (a) Spatial and (b) energy correlation functions for data in Fig. 2 for the HA dataset. Note that significant correlations exist up to ~30 loading map and ~100 energy components. This behavior clearly indicates that linear decomposition methods capture only a part of the relevant physics of the system.

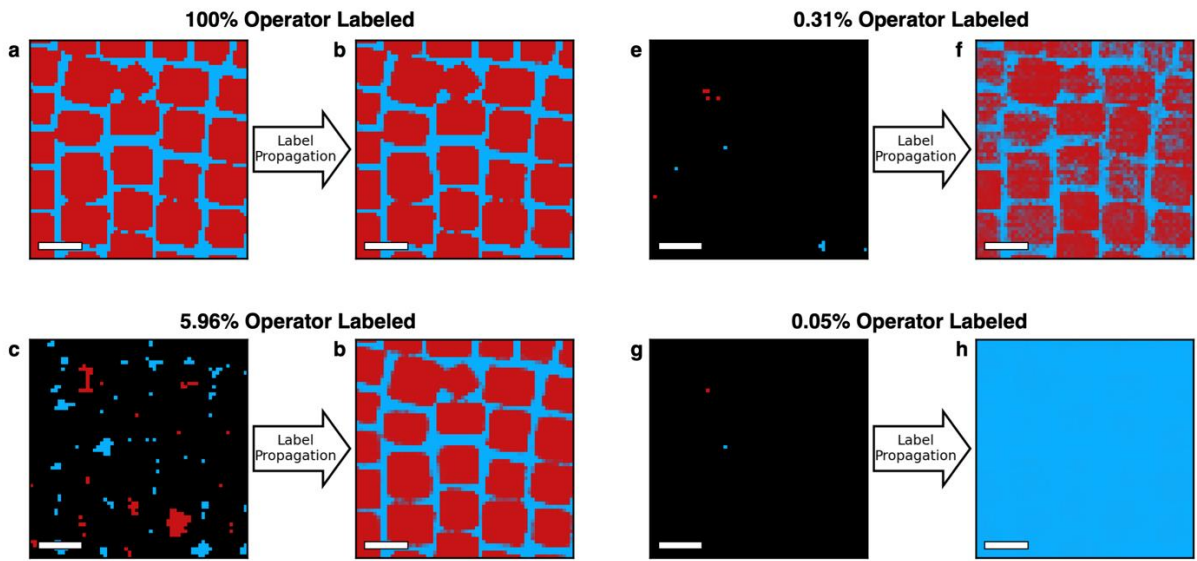

**Figure S2. Influence of Operator-Labeling-Fraction on MLP Label Propagation.** Here, we show the MLP label propagation for a two-label separation in the RA dataset for different amounts of human labeling. (a,b) Show the human labeling (a) and ML propagated labels (b) for when the human operator labels every pixel. (c,d) The same process but where the human labels (5.96%), (e,f) 0.31%, (g,h) 0.05% (two pixels). We can see the MLP label propagation is entirely unaffected between 100% operator-labeling and 5.96% operator-labeling. For extremely low levels of operator-labeling some significant error is introduced into the MLP propagation but the general regions are still effectively identified. In the two-pixel (0.05%) operator-labeling there is not enough data in the training set to produce a meaningful label propagation.

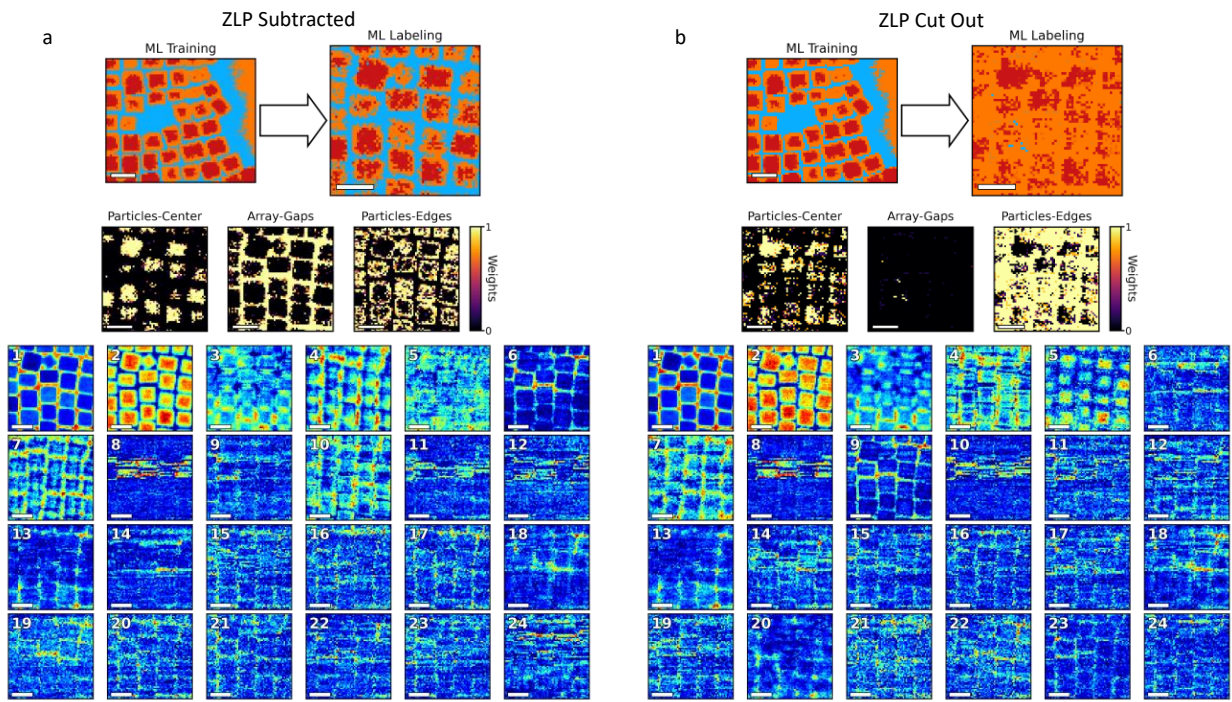

**Figure S3. Transferability of an MLP Network as a Function of Data Preprocessing**. (a) An 3-label MLP network is trained on the HA dataset and applied to the RA dataset with the same mode of background subtraction (power law background fitting). (b) The same process where the RA dataset has the ZLP excluded via cutting out the lower energies of the dataset. We can see clearly if the same mode of background subtraction is used satisfactory results are achieved, but if the background subtraction method is different, all gap areas in the RA dataset are defined as particle edges, indicating that the transferability is sensitive to subtle changes between datasets. This likely connects to the NMF components used to train the MLP initially, and then used to classify the new dataset. The spatial abundance maps of the a 24 component NMF decomposition for the RA dataset under both modes of background subtraction are shown here, and we can see while the first few components are similar, the remaining components are all significantly different, which indicates a reason why a single network could produce such different results on the same dataset.

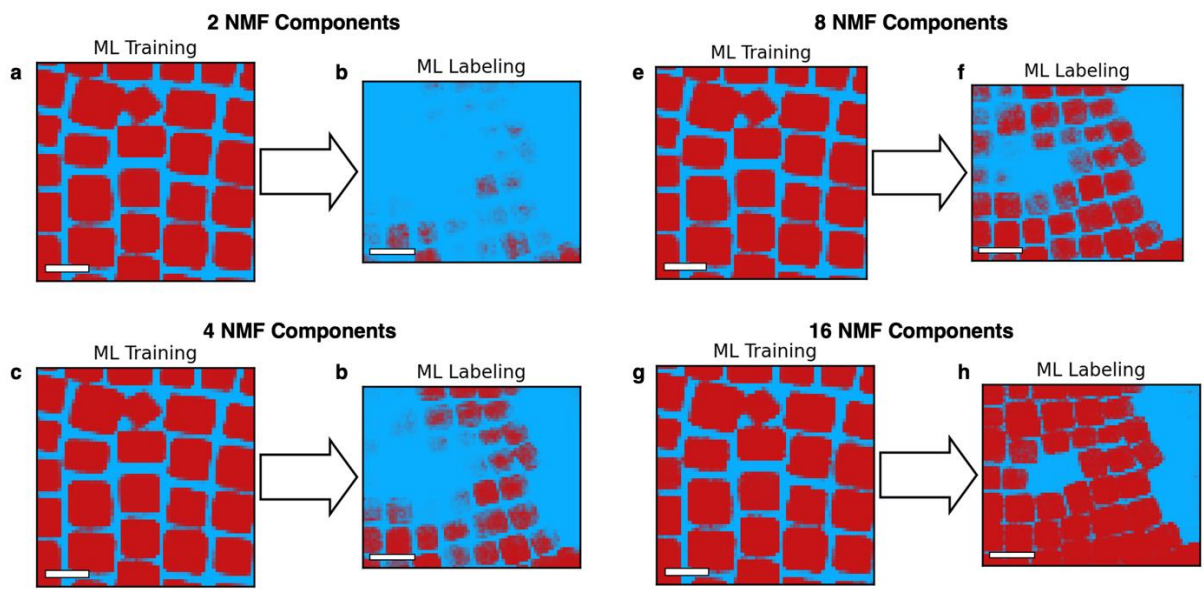

**Figure S4. Transferability of an MLP Network as a function of NMF Components**. (a) A 2-label MLP network is trained on the RA dataset and (b) applied to the HA dataset where 2 NMF components are used for the training and classification. (c,d) Same process with 4 NMF Components. (e,f) 8 NMF Components. (g,h) 16 NMF components. Here, we see that a high amount of NMF components is necessary to achieve robust transferability between different datasets on the same sample.

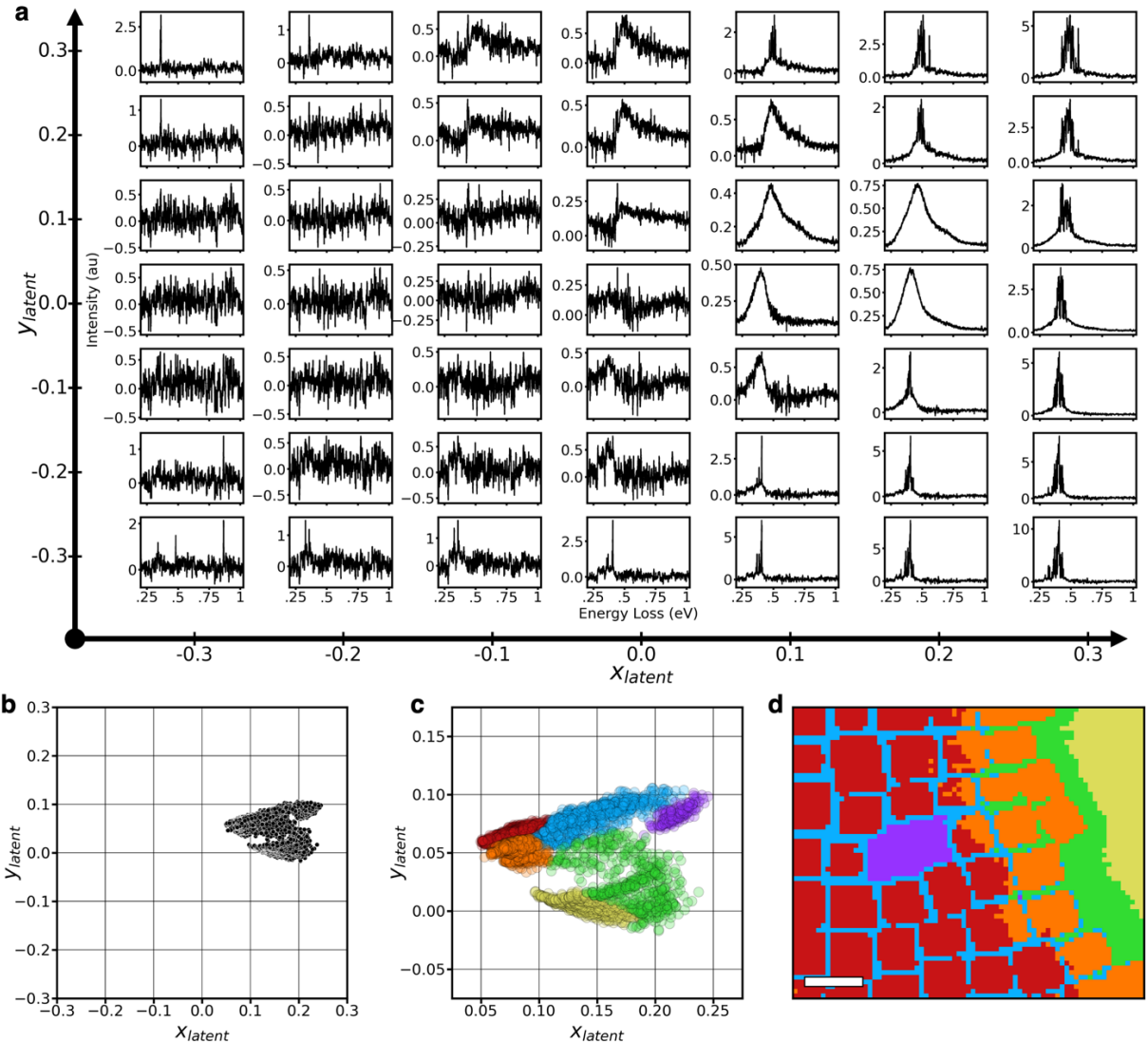

**Figure S5. Decomposition of Hyperspectral Data into Latent Space**. (a) An example of the 2D latent space generated by the autoencoder process, showing a new 2D space made of different spectral features. We can see that many (even most) of the spectral components are just noise (or non physical) however between 0 and 0.1 in the y-dimension and between 0.1 and 0.2 we see spectral components that strongly resemble the plasmonic spectra observed in the dataset (See Fig. 4 and Fig. 7 in the main text as a reference). (b) The coordinates of all 4650 (75x62) spectra in the HA hyperspectral dataset displayed in the 2D latent space (as predicted all coordinates fall in the range of latent space where the components bore a strong physical resemblance to the plasmonic spectrum). (c) With the spectra now assigned to Cartesian coordinates in latent space we can use distance metrics (here we use Gaussian mixture modeling) to cluster coordinates with similar spectral features. (d) The clusters all correspond to spatial pixels in the hyperspectral dataset allowing us to label spectrally distinct regions in the dataset. It is important to note that the coordinates here, do not match quantitatively with the coordinates in the main text, this is because there is a random seeding component to the encoding process that changes what spectral components the latent space values refer to. However, comparing the distribution of the data points

in latent space and the actual cluster assignments to the different encoding shown in the main text demonstrates that the process is robust and repeatable and routinely finds the same behavior with the same pixels.

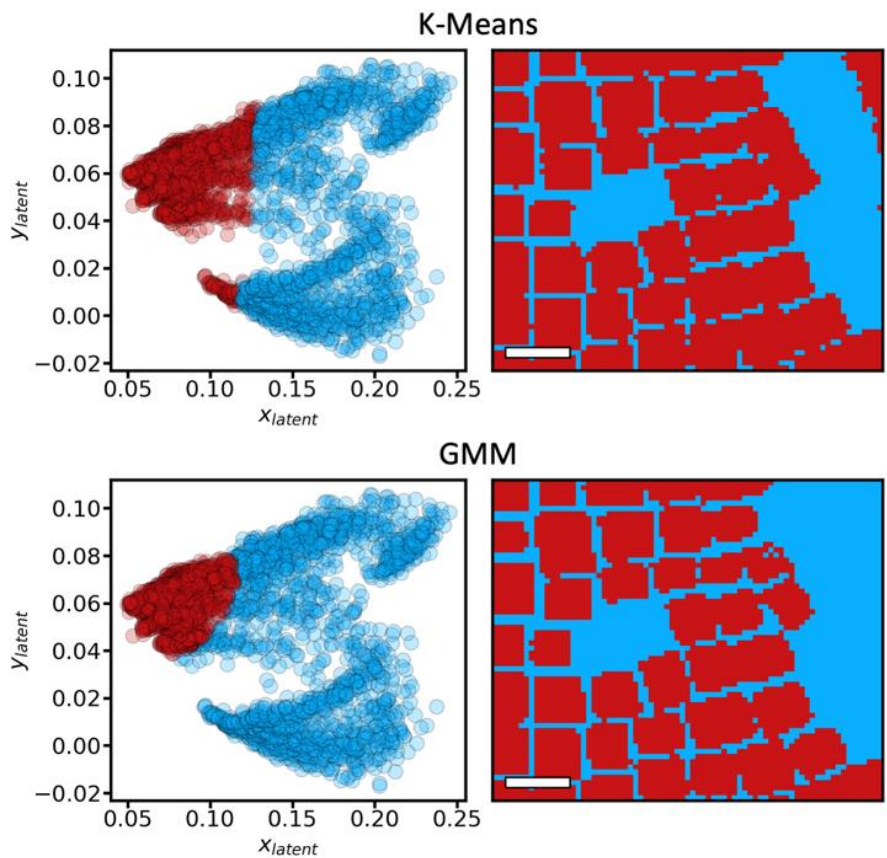

**Figure S6. Clustering Distance Metrics in Latent Space**. The choice of distance metric in the nonlinear manifold generated by the autoencoder is critical. Here, we compare k-means clustering which purely a distance-based metric, to Gaussian mixture modeling (GMM) which attempts to group clusters based off of Gaussian distributions. In the K-means clustering several pixels outside of the self-assembled film are labeled as particles, which is a clearly unphysical result, due to the distance not seeing the separation between the upper and lower cluster of particles. In GMM this separation is recognized for a much more physical separation of the particles and empty space in the HA dataset.

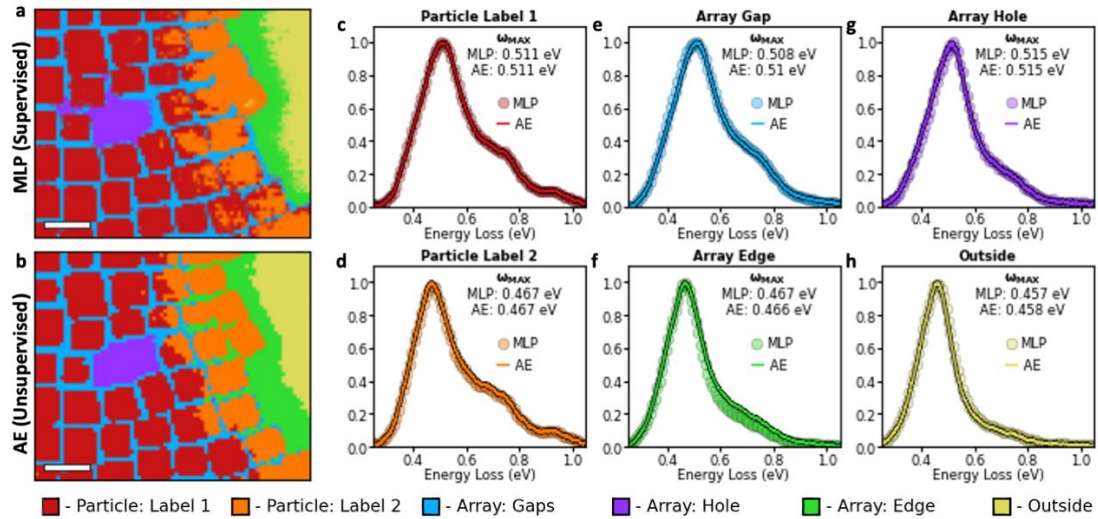

**Figure S7. Isolating Similar Physical Mechanisms with Supervised and Unsupervised Learning.** (a) 6-label operator-assisted MLP label propagation for the HA dataset, now with the 6$^{th}$ label chosen as particles on the edge of the film as opposed to the edges of the particles themselves, (b) 6-GMM classification AE labeling for the HA dataset. Showing similar labels for all aspects other than the two particle labels. (c-h) Comparison of representative spectra between the MLP and AE labeling: (c) Particle Label 1, (d) Particle Label 2, (e) Array Gap, (f) Array Edge, (g) Array Hole, (h) Outside regions. With an operator-selected label that matches the AE result the MLP and AE return almost identical results in terms of the frequency of the ~0.5 eV peak and the intensity of the 0.9 eV peak. We note here that now the 0.9 eV peak looks almost identical in both labels, indicating that the 0.9 eV peak does not couple with the surrounding particles and is identical at the edge of the film as it is in the center (which is not true of the other plasmon modes).